\documentclass{article}
\usepackage{amsmath}
\usepackage{amsfonts}
\usepackage{amsthm}
\usepackage{amssymb}
\usepackage[matrix,arrow,curve]{xy}
\usepackage{array}

\newtheorem{theorem}{Theorem}[section]
\newtheorem{conjecture}[theorem]{Conjecture}
\newtheorem{corollary}[theorem]{Corollary}

\hoffset= - 1.4in

\begin{document}

\newcommand{\loopN}{\begin{picture}(40,50)
%\put(10,10){\oval(20,20)[t]}
\qbezier(0,0)(10,0)(20,10)
\qbezier(20,10)(40,30)(30,40)
\qbezier(30,40)(20,50)(10,40)
\qbezier(20,10)(0,30)(10,40)
\qbezier(40,0)(30,0)(20,10)
%\qbezier(0,#1)(0,.5*#1)(#1,0)
%\qbezier(2*#1,#1)(2*#1,2*#1)(#1,2*#1)
%\qbezier(0,10)(0,5)(10,0)
\end{picture}}

\newcommand{\loopA}{\begin{picture}(40,50)
\qbezier(0,0)(10,0)(20,10)
\qbezier(20,10)(40,30)(30,40)
\qbezier(30,40)(20,50)(10,40)
\qbezier(18,12)(0,30)(10,40)
\qbezier(40,0)(30,0)(22,8)
\end{picture}}

\newcommand{\loopB}{\begin{picture}(40,50)
\qbezier(0,0)(10,0)(18,8)
\qbezier(22,12)(40,30)(30,40)
\qbezier(30,40)(20,50)(10,40)
\qbezier(20,10)(0,30)(10,40)
\qbezier(40,0)(30,0)(20,10)
\end{picture}}

\newcommand{\DblloopA}{\begin{picture}(40,50)
\qbezier(0,0)(10,0)(20,10)
\qbezier(20,10)(40,30)(30,40)
\qbezier(30,40)(20,50)(10,40)
\qbezier(16,14)(0,30)(10,40)
\qbezier(40,0)(30,0)(22,8)
%%%
\qbezier(0,4)(8,4)(17,13)
\qbezier(17,13)(37,30)(27,37)
\qbezier(27,37)(20,45)(13,37)
\qbezier(19,17)(4,30)(13,37)
\qbezier(40,4)(33,4)(24,12)
%%%
\qbezier(21,15)(22,14)(22,14)
\qbezier(18,12)(19,11)(19,11)
\end{picture}}

\newcommand{\R}{\begin{picture}(20,20)
\qbezier(0,0)(5,0)(10,10)
\qbezier(10,10)(15,20)(20,20)
%%%
\qbezier(0,20)(5,20)(9,12)
\qbezier(11,8)(15,0)(20,0)
\end{picture}}

\newcommand{\RInv}{\begin{picture}(20,20)
\qbezier(0,0)(5,0)(9,8)
\qbezier(11,12)(15,20)(20,20)
%%%
\qbezier(0,20)(5,20)(10,10)
\qbezier(10,10)(15,0)(20,0)
\end{picture}}

\title{
\begin{flushright}
\mbox{\normalsize \hfill ITEP/TH-11/14}
\end{flushright}
\vskip 20pt
Khovanov-Rozansky Homologies and Cabling
}
\author{ Danilenko Ivan\\ \\ITEP, Moscow, Russia }
\date{}
\maketitle

\abstract{ In the cabling procedure for HOMFLY polynomials colored HOMFLY polynomials of a knot are obtained from ordinary HOMFLY of the cabled knot with extra twists added. Thus colored polynomials can be seen as relation between HOMFLYs of cabled knot with different twists. In present work we search for relations of such type in Khovanov-Rozansky homologies and investigate, why no generalizations of Khovanov-Rozansky homologies to non-skew-symmetric representations have been constructed. We consider the simplest possible case, i.e. the unknot colored in (11)- and (2)-representations. Naive t-deformation of HOMFLY relation failed to exist in both cases. In case of (11)-representation we have succeeded after a switch to a homological description and that led us to a conjecture about Khovanov-Rozansky homologies of torus knots $T(2,k)$. The (2)-case provides only framing-dependent answers and no simple rule of transformation, such as q,t-shift is seen. We have shown that in our procedure the number of whose nontrivial components with different t-gradings depends on the choice of unknot framing. }

\section{Introduction}
In \cite{Khov} Khovanov introduced a new invariant of knots called Khovanov homologies. The Euler characteristic of these homologies is the $sl(2)$-Jones polynomial, so it may be considered as a deformation of $sl(2)$-Jones polynomial.  In \cite{KhovRoz} the authors generalized this construction to $sl(N)$-Jones polynomial. Their construction is called Khovanov-Rozansky homologies. See \cite{BarNatan,MorDol1,MorDol2,MorDol3} for more details.

Khovanov-Rozansky homologies are uncolored, though there is a construction which colors them in skew-symmetric representations\cite{HaoWu}. This approach is based on MOY-graphs and modification of matrix-factorization construction.

However, for over types of representations such a construction is not yet known. We try to construct a generalized cabling procedure fot Khovanov-Rozansky homologies (see \cite{Adams}). This construction may be used to obtain arbitrary representation coloring on HOMFLY level, so it is a perfect candidate to look for the difference between skew-symmetric representations and others.

Having limited computed examples from \cite{CarqMurf} we investigate a possibility to construct colored unknot. The main idea in this case is to consider different framings. In the HOMFLY case a change of framing acted as multiplication by powers of $q$ and $A = q^N$\cite{AnoAnMor}. Thus we have looked for a homological analogue of this property, an invariance under framing changing up to q,t-shift. In case of (11)-representation we succeeded and moreover provided a conjectural Khovanov-Rozansky polynomial for torus knot $T(2,k)$ which is a consequence of (11)-colored unknot polynomial existence. For (2)-representation it turned out that no colored homology with simple dependence on framing (namely q,t-shift) can be made.

Similar difficulties for non-skew-symmetric representations appeared in the context of superpolynomials. See [???]

\section{Main idea}

Since most of knots do not have a natural choice of framing, an analogue of cabling for Khovanov-Rozansky homologies must not depend on a particular choice of framing. This requirement can be studied for the unknot to figure out the key features of the whole construction (of course, the unknot itself does have a natural framing, because it is a mirror knot, but we should not use this fact to develop a method for arbitrary knot).\\
Let us consider knots with blackboard framing. The framing can be changed with the help of the first Redemeister moves. We consider an unknot $U(k)$ with $k$ Redemeister moves:
$$
\loopA\loopA\cdots\loopA\loopA
$$
For $k<0$ we take an unknot with $|k|$ loops oriented the other way:
$$
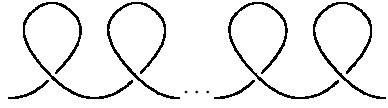
\loopB\loopB\cdots\loopB\loopB
$$
Doing the cabling we obtain
$$
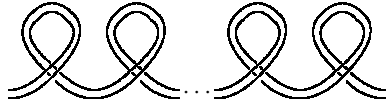
\DblloopA\DblloopA\cdots\DblloopA\DblloopA
$$
Making 2 the first Redemeister moves per each loop we bring it to the standard form of the torus $T(2,2k)$ link:
$$
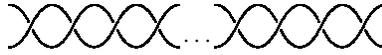
\R\R\R\R\cdots\R\R\R\R
$$
If one adds a twist $\R$ or $\RInv$, one obtains torus knot $T(2,2k+1)$ or $T(2,2k-1)$ respectively.\\
For cabling procedure we need relations between $T(2,2k)$ and $T(2,2k-1)$ (relations for $T(2,2k)$ and $T(2,2k+1)$ are given by the mirror images).

\section{Comparing (11) and (2) for low $k$.}
Let us avoid q-gradings for simplicity and present the dimensions of t-components of Khovanov-Rozansky homologies for several torus knots

\begin{tabular}{c|ccc}
& $t^{-2}$ & $t^{-1}$ & $t^{0}$ \\
\hline
$T(2,2)$ &$N(N-1)$&$0$&$N$ \\
$T(2,1)$ &&&$N$ \\
\hline
$T(2,0)$ &&&$N^2$ \\
$T(2,-1)$ &&&$N$ \\
\end{tabular}

There are two pairs for cablings. In $t=-1$ limit (2)-case corresponds to a sum of two polynomials
\begin{equation}
2 s_2 = Kh^*_{2k} + Kh^*_{2k-1} 
\end{equation}
and (11)-case to their difference
\begin{equation}
2 s_{11} = Kh^*_{2k} - Kh^*_{2k-1} ,
\end{equation}
where $Kh^*_{i}$ is the value of the Khovanov-Rozansky polynomial at the point $t = -1$, $q = 1$.\\
If we look only on the pair $(T(2,-1),T(2,0))$, the most natural thing is to sum them with different coefficients, a priori t-dependent, to obtain (2)- and (11)-case polynomials. However, the pair $(T(2,1),T(2,2))$ makes trouble in the (2)-case. The difference in (11) may cancel two similar terms $N$ in the pair $(T(2,1),T(2,2))$ and provide an expression with a single nonzero t-grading term. The sum in (2) can not do this and thus the polynomial for (2) should have several t-grading terms.

Surely this observation is very naive and may be cured by a modification of the construction. But it clarifies the origin of the problem. As we would see later the number of nontrivial t-grading terms in (2)-case can not be made framing-independent even with a more general construction. 
Also in (11) the cabling is not as trivial as one can think looking at this two pairs. Below we use all the data on $T(2,k)$ from \cite{CarqMurf} and see why difficulties can be avoided for (11) and not for (2).

\section{Polynomial level}

HOMFLY polynomials $H_k$ for torus knots $T(2,k)$ are given by the formula
\begin{equation}
H_k = q^{-k}\cdot s_2 + (-q)^k \cdot s_{11}
\end{equation}
and thus there are relations
\begin{equation}
H_{2k} + q\cdot H_{2k-1} =q^{-2k+1}[2]s_2 \label{H2}
\end{equation}
\begin{equation}
H_{2k}- q^{-1}\cdot H_{2k-1} =-(-q)^{2k-1}[2]s_{11} \label{H11}
\end{equation}
They can be written in the common form
\begin{equation}
\gamma_k\cdot H_{2k} + \tilde\gamma_k\cdot H_{2k-1} = 1,
\end{equation}
where $\gamma_k$ and $\tilde\gamma_k$ are geometric progressions. We have investigated whether there are relations of such form for Khovanov-Rozansky polynomials. The reason why $\gamma_k$ and $\tilde\gamma_k$ are geometric progressions in the HOMFLY case is that $T(2,2k)$ and $T(2,2k-1)$ are cablings of the framed unknot up to $2k$ and $2k-1$ the first Redemeister moves respectively and each brings a multiplier. Assuming this origin is still the only one which prevents $\gamma_k$ and $\tilde\gamma_k$ from being constant we consider $\gamma_k$ and $\tilde\gamma_k$ as geometric progressions. We do not suppose that $\gamma_k$ and $\tilde\gamma_k$ have common ratio since they can be t-deformed. Carqueville and Murfet\cite{CarqMurf} have computed polynomials for $T(2,k)$ up to $k=5$. So there are 5 pairs: $T(2,-5)$ and $T(2,-4)$, $T(2,-3)$ and $T(2,-2)$, $T(2,-1)$ and $T(2,0)$, $T(2,1)$ and $T(2,2)$, $T(2,3)$ and $T(2,4)$. Each 4 pairs restrict all geometric progressions to a finite set of possibilities. Explicit computation shows that for $t=-1$ there are exactly 2 sequences, but \textbf{ for arbitrary $t$ there are none}.\\
This leads to the following statement: the cabling does not appear on polynomial level for Khovanov-Rozansky homologies.

\section{Complexes}

Since HOMFLY polynomials are Euler characteristics of complexes, it is worth looking for homology constructions which lead to proper operations on Euler characteristics.\\
As is widely known for a short exact sequence of complexes
\begin{equation}
\xymatrix{
0\ar@{->}[r]&C^\bullet_0 \ar@{->}[r] & C^\bullet \ar@{->}[r] &C^\bullet_1\ar@{->}[r]&0
} \label{Short}
\end{equation}
the following long exact sequence can be written with the help of the snake lemma
\begin{equation}
\xymatrix{
& \cdots &\ar@{->}[dll] \\
H^i_0 \ar@{->}[r] & H^i \ar@{->}[r] &H^i_1\ar@{->}[dll]\\
H^{i-1}_0 \ar@{->}[r] & H^{i-1} \ar@{->}[r] &H^{i-1}_1\ar@{->}[dll]\\
& \cdots &
}
\end{equation}
Here $H^i_0$, $H^i$ and $H^i_1$ are the $i$th homologies of the corresponding complex. As a consequence for Euler characteristics
\begin{equation}
\chi(C^\bullet_0) - \chi(C^\bullet) + \chi(C^\bullet_1) = 0 \label{Euler}
\end{equation}
holds.\\
The simplest case of (\ref{Short}) is $C^\bullet = C^\bullet_0\oplus C^\bullet_1$ with natural mappings, but that would lead to relations on the polynomial level. Thus if there exists such a sequence, it does not split.\\
Let us present Khovanov-Rozansky homologies $Kh_k$ for torus knots $T(2,k)$\footnote{We have renormalized the answers of Carqueville and Murfet\cite{CarqMurf}: $Kh_k = q^{-kN}\cdot Kh^{CM}_k$ and use the common notation $[k] = \frac{q^{k}-q^{-k}}{q-q^{-1}}$.}:\\

\begin{tabular}{c|ccccccccccc}
 & $t^{-5}$ & $t^{-4}$ & $t^{-3}$ & $t^{-2}$ & $t^{-1}$ & $t^{0}$ & $t^{1}$ & $t^{2}$ & $t^{3}$ & $t^{4}$ & $t^{5}$ \\
\hline
5 &$q^{N+3}[N-1]$&$q^{-N+3}[N-1]$&$q^{N-1}[N-1]$&$q^{-N-1}[N-1]$&$0$&$q^{-N-4}[N]$&&&&& \\
4 &&$q^{3}[N-1][N]$&$q^{N}[N-1]$&$q^{-N}[N-1]$&$0$&$q^{-N-3}[N]$&&&&& \\
3 &&&$q^{N+1}[N-1]$&$q^{-N+1}[N-1]$&$0$&$q^{-N-2}[N]$&&&&& \\
2 &&&&$q[N-1][N]$&$0$&$q^{-N-1}[N]$&&&&& \\
1 &&&&&&$q^{-N}[N]$&&&&& \\
0 &&&&&&$[N]^2$&&&&& \\
-1 &&&&&&$q^{N}[N]$&&&&& \\
-2 &&&&&&$q^{N+1}[N]$&$0$&$q^{-1}[N-1][N]$&&& \\
-3 &&&&&&&&&&& \\
-4 &&&&&&&&&&\lefteqn{\ddots}& \\
-5 &&&&&&&&&&&
\end{tabular}
For $q = 1$ it simplifies:

\begin{tabular}{c|ccccccccccc}
 & $t^{-5}$ & $t^{-4}$ & $t^{-3}$ & $t^{-2}$ & $t^{-1}$ & $t^{0}$ & $t^{1}$ & $t^{2}$ & $t^{3}$ & $t^{4}$ & $t^{5}$ \\
\hline
5 &$N-1$&$N-1$&$N-1$&$N-1$&$0$&$N$&&&&& \\
4 &&$N(N-1)$&$N-1$&$N-1$&$0$&$N$&&&&& \\
3 &&&$N-1$&$N-1$&$0$&$N$&&&&& \\
2 &&&&$N(N-1)$&$0$&$N$&&&&& \\
1 &&&&&&$N$&&&&& \\
0 &&&&&&$N^2$&&&&& \\
-1 &&&&&&$N$&&&&& \\
-2 &&&&&&$N$&$0$&$N(N-1)$&&& \\
-3 &&&&&&$N$&$0$&$N-1$&$N-1$&& \\
-4 &&&&&&$N$&$0$&$N-1$&$N-1$&$N(N-1)$& \\
-5 &&&&&&$N$&$0$&$N-1$&$N-1$&$N-1$&$N-1$
\end{tabular}

It is almost trivial that in one case ``vertical'' lines may cancel each other. That case corresponds to the ``$(11)$-representation'' case, which we discuss below.

\subsection{$(11)$-case}
In this case we have succeeded in the search for such a homology, which is Khovanov-Rozansky analogue of the colored HOMFLY. It has nonzero dimension only in the single t-grading. Unfortunately it has an additional multiplier $[2]$ in its q-dimension and it is difficult to eliminate this multiplier in the homological language. The construction in \cite{HaoWu} is free of such a multiplier.\\

Let us take a pair of Khovanov-Rozansky complexes $Kh_{2k}$ and $Kh_{2k-1}$. For their Euler characteristics to be subtracted with coefficient $q^{-1}$ in (\ref{Euler})-type formula (which should reproduce (\ref{H11})) we need to shift down the $q$-degree of $Kh_{2k-1}$ by $1$ and to place them next to each other in the (\ref{Short})-complex. There are two pairs of inequivalent ways to do it:
\begin{equation}
\xymatrix{
0\ar@{->}[r]&Kh_{2k}^\bullet \ar@{->}[r] & Kh_{2k-1}^\bullet\lbrace-1\rbrace \ar@{->}[r] &C_k^\bullet\ar@{->}[r]&0
}
\end{equation}
\begin{equation}
\xymatrix{
0\ar@{->}[r]&C_k^\bullet\ar@{->}[r]&Kh_{2k}^\bullet \ar@{->}[r] & Kh_{2k-1}^\bullet\lbrace-1\rbrace \ar@{->}[r] &0
}
\end{equation}
and
\begin{equation}
\xymatrix{
0\ar@{->}[r]&Kh_{2k-1}^\bullet\lbrace-1\rbrace \ar@{->}[r]&Kh_{2k}^\bullet \ar@{->}[r] & C_k^\bullet\ar@{->}[r] &0
}
\end{equation}
\begin{equation}
\xymatrix{
0\ar@{->}[r]&C_k^\bullet\ar@{->}[r]&Kh_{2k-1}^\bullet\lbrace-1\rbrace \ar@{->}[r] &Kh_{2k}^\bullet \ar@{->}[r] &0.
}
\end{equation}

Up to redefinition of $C_k^\bullet$ they lead to long exact sequences

\begin{equation}
\xymatrix{
 \cdots\ar@{->}[r] &
H^i_{2k-1} \ar@{->}[r] &H^i_{2k} \ar@{->}[r] &H^i_C\ar@{->}[r]&
H^{i-1}_{2k-1} \ar@{->}[r] &H^{i-1}_{2k} \ar@{->}[r] &H^{i-1}_C\ar@{->}[r]&
\cdots
} \label{FirstCase}
\end{equation}
and
\begin{equation}
\xymatrix{
 \cdots\ar@{->}[r] &
H^i_{2k} \ar@{->}[r] & H^i_{2k-1} \ar@{->}[r] &H^i_C\ar@{->}[r]&
H^{i-1}_{2k} \ar@{->}[r] & H^{i-1}_{2k-1} \ar@{->}[r] &H^{i-1}_C\ar@{->}[r]&
\cdots
} \label{SecondCase}
\end{equation}

\subsubsection{Drawing the sequence (\ref{FirstCase})}
Let us consider the first possible case, the sequence(\ref{FirstCase}). We will see, that in this case we will not obtain proper $C_k^\bullet$s.\\
We consider long exact sequences in low $|k|$ cases:
\begin{itemize}
\item $k=0$
	\begin{equation}
	\xymatrix{
	&0\ar@{->}[dl] && [N]^2\ar@{->}[dl] && 0\ar@{->}[dl]&\\
	0&& \triangle\ar@{->}[dl] && \triangle\ar@{->}[dl] &&0\ar@{->}[dl]\\
	&0\ar@{->}[uu] && q^{N-1}[N]\ar@{->}[uu] && 0\ar@{->}[uu]&\\
	}
	\end{equation}
	The marks $\triangle$ show the place, where nontrivial terms may arise. \\
	Suppose all $C_k^\bullet$ are the same complex up to q- or t-shift. Thus $C_k^\bullet$ have at least one nontrivial homology. If it has two of them, they are next to each other. It can not have more than two nontrivial homology terms.\\
	As we will see later, there is only one nontrivial term.% and to provide an equidistant q-t-shift for $C_k^\bullet$ $$
\item $k=1$
	\begin{equation}
	\xymatrix{
	0&&q[N][N-1]\ar@{->}[dl]&&0\ar@{->}[dl] && q^{-N-1}[N]\ar@{->}[dl] && 0\ar@{->}[dl]&\\
	&\square\ar@{->}[dl] &&0\ar@{->}[dl]&& \diamondsuit\ar@{->}[dl] && \diamondsuit\ar@{->}[dl] &&0\ar@{->}[dl]\\
	0\ar@{->}[uu] &&0\ar@{->}[uu] &&0\ar@{->}[uu] && q^{-N-1}[N]\ar@{->}[uu] && 0\ar@{->}[uu]&\\
	}
	\end{equation}
	At the mark $\square$ there \textbf{must} be a nontrivial term. At the marks $\diamondsuit$ there might be nontrivial terms. With the help of our proposal for $C_k^\bullet$s we see, that $\diamondsuit$s are trivial terms and $C_k^\bullet$s contain only one nontrivial homology.
\item $k=-1$\\
	For this case a problem appears
	\begin{equation}
	\xymatrix{
	0 && q^{N+1}[N]\ar@{->}[dl] && 0\ar@{->}[dl]&&q^{-1}[N][N-1]\ar@{->}[dl] &&0\ar@{->}[dl]&&0\ar@{->}[dl]\\
	& \diamondsuit\ar@{->}[dl] && \diamondsuit\ar@{->}[dl] &&\triangle\ar@{->}[dl]&& \triangle\ar@{->}[dl] && \square\ar@{->}[dl] &\\
	0\ar@{->}[uu] && q^{N+1}[N]\ar@{->}[uu] && 0\ar@{->}[uu]&& q^{N-2}[N-1]\ar@{->}[uu] && q^{-N-2}[N-1]\ar@{->}[uu] &&0\ar@{->}[uu]\\
	}
	\end{equation}
	 In this configuration there must be two nontrivial terms (one at $\square$ and one at one of $\triangle$s). Thus the proposal can not hold.
\end{itemize}

\subsubsection{Drawing the sequence (\ref{SecondCase})}
In this case we have found such $C_k^\bullet$s, which differ only by a q,t-shift. Therefore they may be considered as a generalization of $(11)$-colored HOMFLY for unknot.

Most of the comments for the diagrams are the same as for the previous sequence, so we skip them.
\begin{itemize}
\item $k=0$
	\begin{equation}
	\xymatrix{
	&0\ar@{->}[dd] && [N]^2\ar@{->}[dd] && 0\ar@{->}[dd]&\\
	0&& \triangle\ar@{->}[ul] && \triangle\ar@{->}[ul] &&0\ar@{->}[ul]\\
	&0\ar@{->}[ul] && q^{N-1}[N]\ar@{->}[ul] && 0\ar@{->}[ul]&\\
	}
	\end{equation}
\item $k=1$
	\begin{equation}
	\xymatrix{
	0\ar@{->}[dd] &&q[N][N-1]\ar@{->}[dd]&&0\ar@{->}[dd] && q^{-N-1}[N]\ar@{->}[dd] && 0\ar@{->}[dd]&\\
	&0\ar@{->}[ul] &&\square\ar@{->}[ul]&& \diamondsuit\ar@{->}[ul] && \diamondsuit\ar@{->}[ul] &&0\ar@{->}[ul]\\
	0 &&0\ar@{->}[ul] &&0\ar@{->}[ul] && q^{-N-1}[N]\ar@{->}[ul] && 0\ar@{->}[ul]&\\
	}
	\end{equation}
	If one of $\diamondsuit$ is nontrivial, the other is also nontrivial. Thus they vanish and $C_k^\bullet$s have a single nontrivial homology.
\item $k=-1$
	\begin{equation}
	\xymatrix{
	0\ar@{->}[dd] && q^{N+1}[N]\ar@{->}[dd] && 0\ar@{->}[dd]&&q^{-1}[N][N-1]\ar@{->}[dd] &&0\ar@{->}[dd]&&0\ar@{->}[dd]\\
	& \diamondsuit\ar@{->}[ul] && \diamondsuit\ar@{->}[ul] &&\diamondsuit\ar@{->}[ul]&& \square\ar@{->}[ul] && 0\ar@{->}[ul] &\\
	0 && q^{N+1}[N]\ar@{->}[ul] && 0\ar@{->}[ul]&& q^{N-2}[N-1]\ar@{->}[ul] && q^{-N-2}[N-1]\ar@{->}[ul] &&0\ar@{->}[ul]\\
	}
	\end{equation}
	Since $C_k^\bullet$s have a single nontrivial homology, the homologies at $\diamondsuit$s vanish.
\end{itemize}

It is natural to sum up this experimental data in a conjecture:
\begin{conjecture}
$C_k^\bullet$ has a single nontrivial homology term with $q$-dimension $q^{2k-1}[N][N-1]$ and for $C_{k+1}^\bullet$ this term is shifted by two t-gradings to the left for all $k$.
\label{conjecture_main}
\end{conjecture}

In other words the Poincar\'e polynomial of $C_k^\bullet$ is $P^k_C(q,t) = t^{-2k+n_0}q^{2k-1}[N][N-1]$, where $n_0$ defines the t-grading position of this term when $k=0$ (this depends on the definition of $C_k^\bullet$).

\subsubsection{Conjectures from (\ref{SecondCase})}
The existence of such a long exact sequence provides an opportunity to make predictions from Conjecture(\ref{conjecture_main}).\\
Let us denote the q-dimension of the Khovanov-Rozansky homologies $KhR$ component with $l$th t-grading as $KhR^l$ and the Khovanov-Rozansky homologies for torus $T(2,k)$ knot or link as $KhR_{k}$. Then the following corollary holds:

\begin{corollary}
For all t-gradings $l$ except $-k$ and $-k+1$ the following condition holds: $q\cdot KhR_k^l = KhR_{k-1}^l$.
\end{corollary}

This is true because either $KhR_k$ and $KhR_{k-1}$ or $KhR_{-k}$ and $KhR_{-k+1}$ are connected by the presented long sequence and $C_k^\bullet$ has a single nonzero term at the fragment
\begin{equation}
\xymatrix{
 \cdots\ar@{->}[r] & 0\ar@{->}[r] &
H^{-2k'+1}_{2k'} \ar@{->}[r] & H^{-2k'+1}_{2k'-1} \ar@{->}[r] &H^{-2k'+1}_C\ar@{->}[r]&
H^{-2k'}_{2k'} \ar@{->}[r] & H^{-2k'}_{2k'-1} \ar@{->}[r] &0\ar@{->}[r]&
\cdots }. \label{frag}
\end{equation}

The structure of that fragment is also similar for different $k'$s. Let us introduce another conjecture
\begin{conjecture}
For $k' > 0$ the fragment(\ref{frag}) is as follows
\begin{equation}
\xymatrix{
	\cdots && q^{2k'+1}[N][N-1] \ar@{->}[dd] && \triangledown \ar@{->}[dd] &&\\
	&0\ar@{->}[ul] &&q^{2k'+1}[N][N-1]\ar@{->}[ul]&& 0\ar@{->}[ul] & \\
	&&0\ar@{->}[ul] && \triangledown \ar@{->}[ul] && \cdots \ar@{->}[ul]
}
\end{equation}
where $ \triangledown $s mark equal terms.
For $k' < 0$ the fragment(\ref{frag}) is as follows
\begin{equation}
	\xymatrix{
	\cdots&&q^{2k'+1}[N][N-1]\ar@{->}[dd] &&0\ar@{->}[dd]&&\\
	&0\ar@{->}[ul]&& q^{2k'-1}[N][N-1]\ar@{->}[ul] && 0\ar@{->}[ul] &\\
	&& q^{N+2k'}[N-1]\ar@{->}[ul] && q^{-N+2k'}[N-1]\ar@{->}[ul] &&\cdots\ar@{->}[ul]\\
	}
	\end{equation}
\end{conjecture}

From this conjectures one can make a prediction for $T(2,k)$ knots and links, because $T(2,k+1)$ is automatically reconstructed from $T(2,k)$:

\begin{corollary}
For $k'>0$
\begin{align}
KhR_{2k'}^{l} &= 0 \;\; for \; l>0\; or \; l<-2k'; \\
KhR_{2k'}^{0} &= q^{-N-2k'+1}[N]; \\
KhR_{2k'}^{-1} &= 0;\\
KhR_{2k'}^{-2l-1} &= q^{N+4l-2k'}[N-1] \;\; for \; 1\leq l < k'; \\
KhR_{2k'}^{-2l} &= q^{-N+4l-2k'}[N-1] \;\; for \; 1\leq l < k'; \\
KhR_{2k'}^{-2k'} &= q^{2k'-1}[N][N-1].
\end{align}
and
\begin{align}
KhR_{2k'+1}^{l} &= 0 \;\; for \; l>0\; or \; l<-2k'+1; \\
KhR_{2k'+1}^{0} &= q^{-N-2k'}[N]; \\
KhR_{2k'+1}^{-1} &= 0;\\
KhR_{2k'+1}^{-2l-1} &= q^{N+4l-2k'-1}[N-1] \;\; for \; 1\leq l \leq k'; \\
KhR_{2k'+1}^{-2l} &= q^{-N+4l-2k'-1}[N-1] \;\; for \; 1\leq l < k'.
\end{align}
\end{corollary}

This agrees with Carqueville and Murfet\cite{CarqMurf} for all available data, even with $T(2,6)$, which is calculated only for $N=3$ and $N=4$. Also the answers for torus $T(2,k)$ coincide with \cite{AnoMor} up to a change of variables $T=t^{-1}$ and a common q-shift.

\subsection{$(2)$-case}

In this case no simple analogue of colored HOMFLY is found. The number of nonvanishing t-grading terms is growing with $k$, so the connection between the answers for different cablings can not be simple q,t-shift. It may be changed to something more sophisticated. Also the projector on (2) may no longer be constructed from twists (R-matrixes) in Khovanov case. However there is not enough data to say precisely how.

The sequences which reproduce familiar formulae on Euler-characteristic level for low $k$ are as follows (the vertical lines are arrows too and must be chosen all up or all down. Unfortunately, both of this cases have the same problem)

\begin{itemize}
\item $k=0$
	\begin{equation}
	\xymatrix{
	0 & [N]^2\ar@{-->}[ddl] & 0\ar@{-->}[ddl]\\
	0\ar@{-}[u]\ar@{-}[d]& \square\ar@{-}[u]\ar@{-}[d] &0\ar@{-}[u]\ar@{-}[d]\\
	0 & q^{N+1}[N]\ar@{-->}[uul] & 0\ar@{-->}[uul]\\
	}
	\end{equation}
\item $k=1$
	\begin{equation}
	\xymatrix{
	0 &q[N][N-1]\ar@{-->}[ddl] & 0\ar@{-->}[ddl] & q^{-N-1}[N]\ar@{-->}[ddl] & 0\ar@{-->}[ddl]\\
	0\ar@{-}[u]\ar@{-}[d] &\square\ar@{-}[u]\ar@{-}[d]& 0\ar@{-}[u]\ar@{-}[d] & \square\ar@{-}[u]\ar@{-}[d] &0\ar@{-}[u]\ar@{-}[d]\\
	0 &0\ar@{-->}[uul] &0\ar@{-->}[uul] & q^{-N+1}[N]\ar@{-->}[uul] & 0\ar@{-->}[uul]\\
	}
	\end{equation}
\item $k=-1$
	\begin{equation}
	\xymatrix{
	0 & q^{N+1}[N]\ar@{-->}[ddl] & 0\ar@{-->}[ddl] &q^{N-1}[N][N-1]\ar@{-->}[ddl] &0\ar@{-->}[ddl] &0\ar@{-->}[ddl] \\
	0\ar@{-}[u]\ar@{-}[d] & \square\ar@{-}[u]\ar@{-}[d] &0\ar@{-}[u]\ar@{-}[d] & \square\ar@{-}[u]\ar@{-}[d] & \square\ar@{-}[u]\ar@{-}[d] & 0\ar@{-}[u]\ar@{-}[d] \\
	0 & q^{N+1}[N]\ar@{-->}[uul] & 0\ar@{-->}[uul] & q^{N-2}[N-1]\ar@{-->}[uul] & q^{-N-2}[N-1]\ar@{-->}[uul] &0\ar@{-->}[uul] \\
	}
	\end{equation}
\item $k=2$
	\begin{equation}
	\xymatrix{
	0 &q^3[N][N-1]\ar@{-->}[ddl] & q^{N}[N-1]\ar@{-->}[ddl] & q^{-N}[N-1]\ar@{-->}[ddl] & 0\ar@{-->}[ddl] & q^{-N-3}[N]\ar@{-->}[ddl] & 0\ar@{-->}[ddl]\\
	0\ar@{-}[u]\ar@{-}[d] & \square\ar@{-}[u]\ar@{-}[d] & \square\ar@{-}[u]\ar@{-}[d] &\square\ar@{-}[u]\ar@{-}[d]& 0\ar@{-}[u]\ar@{-}[d] & \square\ar@{-}[u]\ar@{-}[d] &0\ar@{-}[u]\ar@{-}[d]\\
	0 &0\ar@{-->}[uul] & q^{N+2}[N-1]\ar@{-->}[uul] & q^{-N+2}[N-1]\ar@{-->}[uul] &0\ar@{-->}[uul] & q^{-N+1}[N]\ar@{-->}[uul] & 0\ar@{-->}[uul]\\
	}
	\end{equation}
\end{itemize}
As it is in the previous section, $\square $ denotes the place, where a nontrivial term \textbf{must} be.\\
As we see, the number of nontrivial homology terms grows. It is unclear yet if there is any general rule for $C_k^\bullet$, but it can not be simple q,t-grading shift as in (11)-case.

\section{Conclusion}

We have investigated a possibility to obtain an analogue of the cabling procedure for Khovsnov-Rozansky homologies and have found a difference between (11)- and (2)-representation cases. Having limited experimental data with computed Khovanov-Rozansky polynomials from \cite{CarqMurf} we have investigated the colored unknot. Though cabling fails on polynomial level, we have made an attempt to construct it on homological level. We have used only q-dimensions and thus we had to guess possible morphisms. However these morphisms look really natural.

For (11)-representation the construction succeeded for available data. The conjecture for the behavior of the sequences appearing in this formalism was presented. The conjecture also provided a method to obtain Khovanov-Rozansky homologies for the arbitrary $T(2,k)$.

The (2)-colored unknot appeared not to have a colored Khovanov-Rozansky polynomial independent from unknot framing (up to q,t-shift). The dependence on framing may be more sophisticated and should be examined in the future.

\section*{Acknowledgements}

The author is grateful to Alexei Morozov and Alexandra Anokhina for discussions and to Yegor Zenkevich and Gleb Aminov for helpful remarks.\\
Our work is partly supported by RFBR grants 13-02-00478, 14-02-31446-mol-a, NSh-1500.2014.2 and by the common grant 14-01-92691-Ind-a.

\end{document}